\documentstyle[spie]{article}
\input{psfig.sty}

\title{Observations of Mira stars with the IOTA/FLUOR interferometer and comparison with
Mira star models}

\author{K.-H. Hofmann\supit{a}, U. Beckmann\supit{a}, T. Bl\"ocker\supit{a}, V. Coude du Foresto\supit{b}, M. Lacasse\supit{c}, \\
R. Millan-Gabet\supit{c}, S. Morel\supit{c},
B. Pras \supit{c}, C. Ruilier\supit{b}, D. Schertl\supit{a}, M. Scholz \supit{d}, V. Shenavrin\supit{e}, \\
W. Traub\supit{c}, G. Weigelt\supit{a}, M. Wittkowski\supit{a}, B. Yudin\supit{e}
\skiplinehalf
\supit{a}MPI f\"ur Radioastronomie, Auf dem Huegel 69, 53121 Bonn, Germany
\\
\supit{b}Observatoire de Paris-Meudon, 5 place Jules Janssen, \\ 92195 Meudon Cedex, France
\\
\supit{c}Harvard-Smithsonian Center for Astrophysics, 60 Garden Street, \\ Cambridge, Massachusetts 02138, USA
\\
\supit{d}Institut f\"ur Theoretische Astrophysik der Universitaet Heidelberg,Tiergartenstrasse 15, \\ 69121 Heidelberg, Germany
\\
\supit{e} Sternberg Astronomical Institute, Universitetskii pr. 13, \\
119899 Moscow, Russia.
}

\begin{document}
  \maketitle

\begin{abstract}
We present K-band observations of five Mira stars with the IOTA interferometer.
The interferograms were obtained with the FLUOR fiber optics beam combiner which
provides high-accuracy
visibility measurements in spite of time-variable atmospheric conditions.
For the Mira stars X~Oph, R~Aql, RU~Her, R~Ser, and V~CrB we derived the
uniform-disk diameters 11.7~mas, 10.9~mas, 8.4~mas, 8.1~mas, and 7.9~mas
($\pm$0.3 mas), respectively.
Simultaneous photometric observations
yielded the bolometric fluxes. The derived angular Rosseland radii
and the bolometric fluxes allowed the determination of effective
temperatures. For instance, the effective temperature of R~Aql was determined
to be 3072~K\,$\pm$161~K. A Rosseland radius for R~Aql of 250\,R$_{\odot}$\,$\pm$63\,R$_{\odot}$
was derived from the angular Rosseland radius of 5.5\,mas\,$\pm$0.2\,mas and the
HIPPARCOS parallax of 4.73 mas\,$\pm$1.19\,mas. The observations were compared with
theoretical Mira star models$^{1,2}$ (D/P model Rosseland radius\,=\,255\,R$_{\odot}$;
measured R~Aql Rosseland radius\,=\,250\,R$_{\odot}$\,$\pm$63\,R$_{\odot}$).
%of Bessel, Scholz and Wood (1996) and Hofmann, Scholz and Wood (1998).
\end{abstract}


\section{INTRODUCTION}
The resolution of large optical telescopes and interferometers is high enough
to resolve the stellar disk of nearby M giant stars, to reveal photospheric asymmetries
and surface structures, and to study the dependence of the diameter on
the wavelength, variability phase, and cycle.
Previous speckle or long-baseline interferometry observations were, for example,
reported in Refs. 3-10.
% (Bonneau \& Labeyrie 1973; Karovska et al. 1991;
% Quirrenbach et al. 1992; Haniff et al. 1995; Weigelt et al. 1996, van Belle et al. 1996,
% Perrin et al. 1999; Hofmann et al. 2000).
Theoretical studies (e.g. Refs. 1-2 and 11-13)
% Watanabe \& Kodaira 1979; Scholz 1985; Bessell et al. 1989;
% Bessell et al. 1996=BSW96)
show that
accurate monochromatic diameter measurements can significantly improve our understanding
of M giant atmospheres.
%The current generation of interferometers provides
%the astronomical community with approx. an order of magnitude increase in spatial
%resolution over large single-dish telecope speckle observations.
%For example,
With the IOTA interferometer
a resolution of $\sim$\,9 mas can be achieved with its largest
baseline of 38~m in the K-band.
The IOTA interferometer is located at the Smithsonian Institution's Whipple Observatory on Mount Hopkins
in Arizona. A detailed description of IOTA can be found in Refs. 14 and 15.
% Carleton et al. (1994) and
% Traub et al. (1998).
IOTA can be operated in the K-band with the FLUOR$^{16}$ fiber optics beam combiner.
% Coude de Foresto et al. 1997).
This beam combiner provides high-accuracy visibility measurements
in spite of time-variable atmospheric conditions. The single-mode fibers in the beam combiner
spatially filter the wavefronts corrugated by atmospheric turbulence (see Refs. 16 and 17).
% 17: Perrin et al. 1998)

\section{OBSERVATIONS}
\begin{figure}
\begin{center}
\psfig{figure=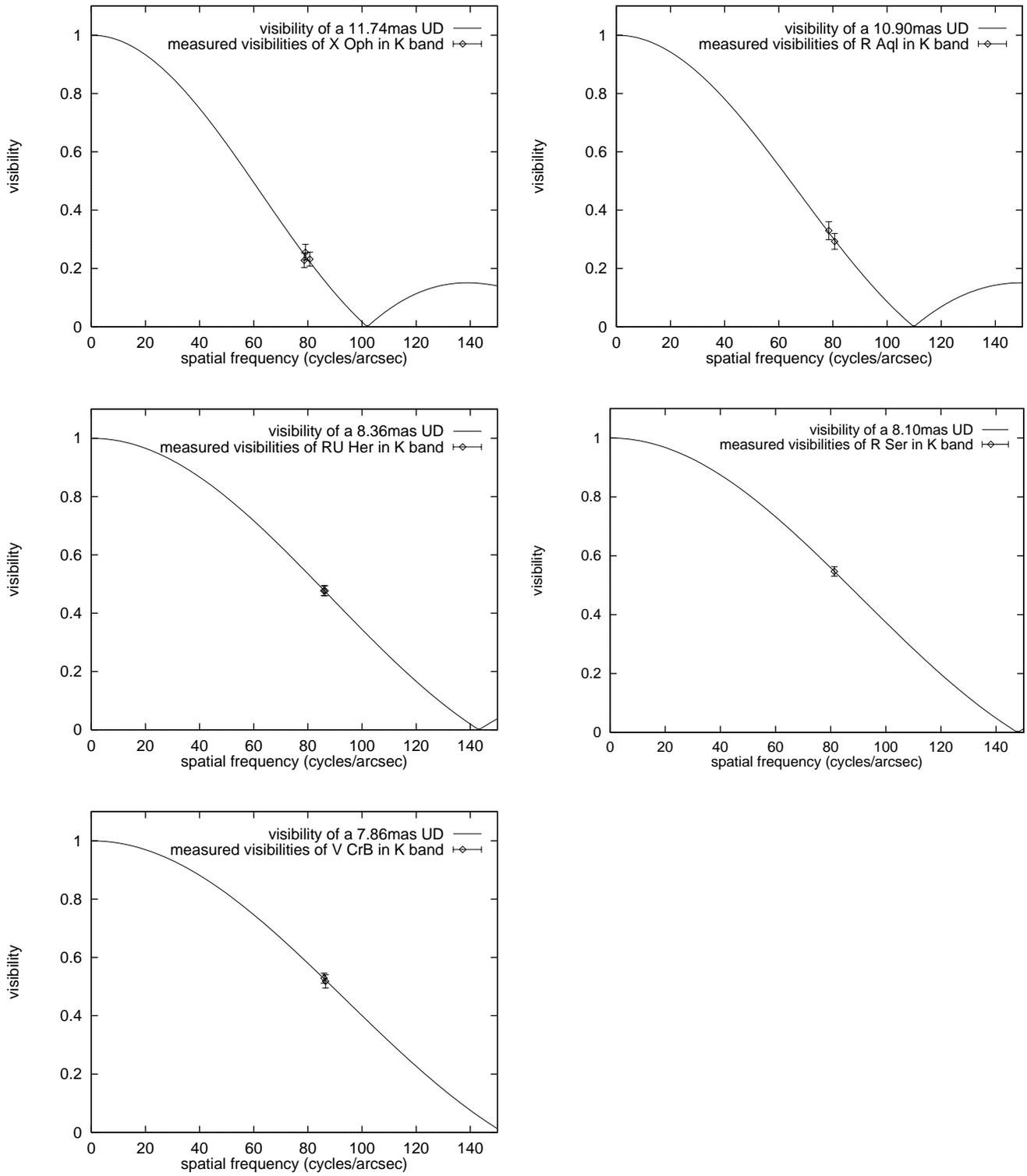,height=20cm}
\caption[Vis]{ \label{fig:Vis} Uniform-disk fits (X~Oph, R~Aql, RU~Her, R~Ser, and V~CrB).}
\end{center}
\end{figure}

The five Miras X~Oph, R~Aql, RU~Her, R~Ser, V~CrB were observed with the IOTA
interferometer on May 16, 17 and 18, 1999. The observations
were carried out with the fiber optics beam combiner FLUOR in the K-band and with
38~m baseline.
The interferograms are scanned by the delay line during the coherence time of the atmosphere.
The OPD length of the scan is $\sim$\,100\,$\mu$m.
Approximately 100 scans per baseline were recorded.
Several reference stars (Table~\ref{tab:obs}) were observed for the calibration of the observations (see Ref. 17 for
more details). The diameters of the reference stars were derived from the scale
of stellar diameters at K-magnitude\,=\,0 for giants by Dyck et al.$^{18}$.
The fringe visibility of the reference stars was 64\%\,-\,94\%. Fig.~\ref{fig:Vis}
shows the obtained visibility functions of the five Mira stars together with uniform-disk fits.
The errors of the derived
Mira star diameters are 1-3\%.

\begin{table} [ht]
\caption[]{Observed data.}
\label{tab:obs}
\begin{center}
{\small
\begin{tabular}{|c|c|c|c|c|c|c|c|c|}
\hline
\rule[-1ex]{0pt}{3.5ex} Star & spectral type & $P$ & Date & $\Phi_{\rm vis}$ & $B_{\rm p}$
       & $V$ & $\Theta_{\rm UD}$ & reference stars \\
\rule[-1ex]{0pt}{3.5ex}      &               &  [days] &  &   &  [m] &  & [mas] & \\
\hline
\rule[-1ex]{0pt}{3.5ex} X Oph & M5e-M9e & 328 & 99 May 17 & 0.71 & 35.47 & 0.2317$\pm$0.024 & 11.74$\pm$0.30 & HIP 86742 \\
\rule[-1ex]{0pt}{3.5ex}       &         &     & 99 May 18 &      & 34.75 & 0.2554$\pm$0.027 &  & HIP 98337 \\
\rule[-1ex]{0pt}{3.5ex}       &         &     & 99 May 18 &      & 34.57 & 0.2279$\pm$0.025 &  & HIP 98438 \\
\rule[-1ex]{0pt}{3.5ex}       &         &     &           &      &       &                  &  & HIP 97278 \\ 
\rule[-1ex]{0pt}{3.5ex}       &         &     &           &      &       &                  &  & HIP 97278 \\ 
\rule[-1ex]{0pt}{3.5ex} R Aql & M5e-M9e & 284 & 99 May 17 & 0.17 & 35.42 & 0.2927$\pm$0.027 & 10.90$\pm$0.33 & HIP 86742 \\
\rule[-1ex]{0pt}{3.5ex}       &         &     & 99 May 18 &      & 34.48 & 0.3295$\pm$0.031 &  & HIP 98337 \\
\rule[-1ex]{0pt}{3.5ex}       &         &     &           &      &       &                  &  & HIP 98438 \\
\rule[-1ex]{0pt}{3.5ex}       &         &     &           &      &       &                  &  & HIP 97278 \\
\rule[-1ex]{0pt}{3.5ex} RU Her & M6e-M9 & 484 & 99 May 17 & 0.07 & 37.95 & 0.4768$\pm$0.017 & 8.36$\pm$0.20 & HIP 71053 \\
\rule[-1ex]{0pt}{3.5ex}        &        &     &           &      & 37.73 & 0.4769$\pm$0.017 & & HIP 78159 \\
\rule[-1ex]{0pt}{3.5ex} R Ser & M5e-M9e & 356 & 99 May 18 & 0.28 & 35.74 & 0.5467$\pm$0.016 & 8.10$\pm$0.20 & HIP 61658 \\
\rule[-1ex]{0pt}{3.5ex}       &            &     &           &      &       &                  &  & HIP 75530 \\
\rule[-1ex]{0pt}{3.5ex}       &            &     &           &      &       &                  &  & HIP 85934 \\
\rule[-1ex]{0pt}{3.5ex} V CrB & C6,2e(N2e) & 357 & 99 May 16 & 0.07 & 37.78 & 0.5288$\pm$0.017 & 7.86$\pm$0.24 & HIP 73555 \\
\rule[-1ex]{0pt}{3.5ex}       &           &      &           &      & 38.02 & 0.5180$\pm$0.023 &  & HIP 81833 \\
\hline
\end{tabular}
}
\end{center}
\end{table}

In Table~\ref{tab:obs} the calibrated visibilities and the derived uniform-disk
diameters of the five Miras are listed, together with observational
parameters (spectral type, variability period $P$, date of observation, variability phase $\Phi_{vis}$,
projected baseline length $B_{\rm p}$, calibrated visibilities $V$, derived uniform-disk diameters $\Theta_{\rm UD}$,
and reference stars).
\section{COMPARISON OF THE OBSERVATIONS WITH MIRA STAR MODELS}
\begin{figure}
\begin{center}
\psfig{figure=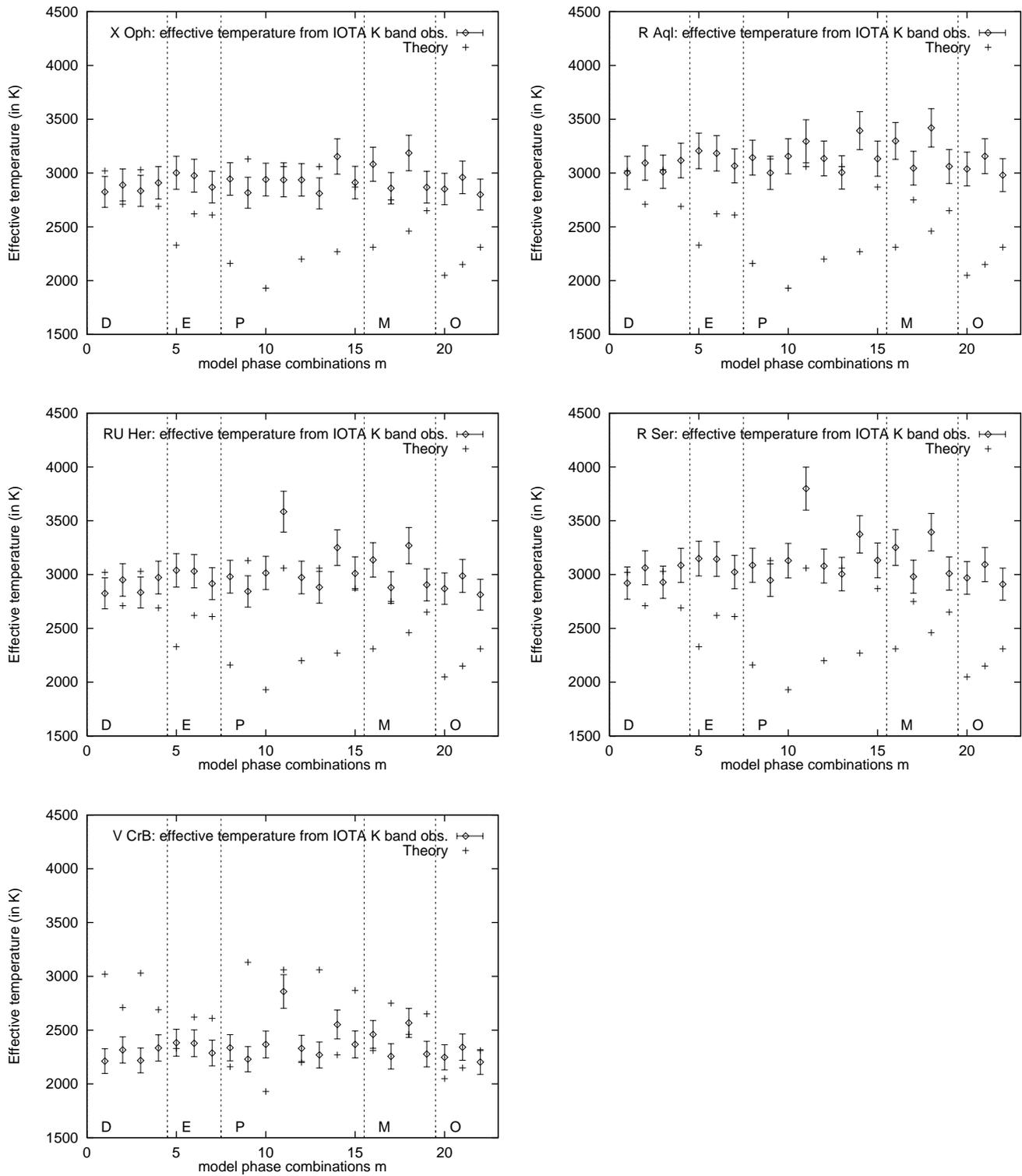,height=20cm}
\caption[Teff]{ \label{fig:Teff} Comparison of measured effective temperatures
of the 5 observed Mira stars and the theoretical model effective temperatures (see text).
Table \ref{tab:link} shows the link between the abscissa values and the models and their
phases.
}
\end{center}
\end{figure}

In this section we derive
angular diameters
from the measured visibilities by fitting
different theoretical center-to-limb intensity variations
(hereafter CLV) of different Mira star models (Bessel, Scholz and Wood 1996 = BSW96$^1$,
Hofmann, Scholz and Wood 1998 = HSW98$^2$).
From these angular diameters
and the bolometric flux, we derive effective temperatures.
For R~Aql
a HIPPARCOS parallax is available which allows us to
determine linear radii.
The comparison of these measured stellar parameters with theoretical ones
indicate whether any of the models are a fair representation of the
observed Mira stars.
All Mira star models used in this paper are from BSW96 (D and E series)
and from HSW98 (P, M and O series). They were developed as possible representations
of the prototype Mira variable o Ceti, and hence have periods $P$
very close to the 332 day period of this star; they differ in pulsation mode, assumed mass $M$ and
assumed luminosity $L$; and the BSW96 models differ from the (more advanced) HSW98 models
with respect to the pulsation modelling technique.
The five models represent stars pulsating in the fundamental mode ($f$; D, P and M models) or
in the first-overtone mode ($o$; E and O models).
Table \ref{tab:prop} lists the properties of these Mira model series ($R_{\rm p}$ = Rosseland radius of the
non-pulsating parent star of the Mira variable
= distance from the "parent star's" center, at which the Rosseland optical depth $\tau_{\rm Ross}$ equals
unity, see BSW96 and HSW98;
$T_{\rm eff} \propto (L/R_{\rm p}^2)^{1/4}$ = effective
temperature). Table \ref{tab:link} provides the link between the 22 abscissa values (model-phase combinations m)
in Figs.~\ref{fig:Teff} and \ref{fig:linradii},
and the models, and it additionally lists the variability phase, relative Rosseland
and stellar K-band filter radius, and the effective temperature.
We compare predictions of these models at different phases and cycles with
our observations.

\begin{table} [ht]
\caption[]{Properties of Mira model series$^{1,2}$ (see text)}
\label{tab:prop}
\begin{center}
{\small
\begin{tabular}{|c|c|c|c|c|c|c|}
\hline
\rule[-1ex]{0pt}{3.5ex} Series & Mode & $P$(days) & $M/M_{\odot}$ & $L/L_{\odot}$
       & $R_{\rm p}/R_{\odot}$ & $T_{\rm eff}$/K \\
\hline
\rule[-1ex]{0pt}{3.5ex} D & f & 330 & 1.0 & 3470 & 236 & 2900  \\
\rule[-1ex]{0pt}{3.5ex} E & o & 328 & 1.0 & 6310 & 366 & 2700  \\
\rule[-1ex]{0pt}{3.5ex} P & f & 332 & 1.0 & 3470 & 241 & 2860  \\
\rule[-1ex]{0pt}{3.5ex} M & f & 332 & 1.2 & 3470 & 260 & 2750  \\
\rule[-1ex]{0pt}{3.5ex} O & o & 320 & 2.0 & 5830 & 503 & 2250  \\
\hline
\end{tabular}
}
\end{center}
\end{table}

\begin{table} [ht]
\caption[]{Link between the 22 abscissa values (model-phase combinations m)
in Figs.~\ref{fig:Teff} and \ref{fig:linradii}, and the models.
The variability phase
$\Phi_{\rm vis}$, the Rosseland radius $R$ and the K-band radius $R_{\rm K}$
in units
of the parent star radius $R_p$, and the effective temperature $T_{\rm eff}(R)$
associated to the Rosseland radius are additionally given.}
\label{tab:link}
\begin{center}
{\small
\begin{tabular}{|l|l|l|l|l|l|}
\hline
\rule[-1ex]{0pt}{3.5ex} Model & $\Phi_{\rm vis}$ & $R/R_{\rm p}$ & $R_{\rm K}/R_{\rm p}$ & $T_{\rm eff}(R)$ & m \\
\hline
\rule[-1ex]{0pt}{3.5ex} D27520 & 1+0.0  &  1.04 & 1.02  & 3020 & 1 \\
\rule[-1ex]{0pt}{3.5ex} D27760 & 1+0.5  &  0.91 & 0.90  & 2710 & 2 \\
\rule[-1ex]{0pt}{3.5ex} D28760 & 2+0.0  &  1.04 & 1.02  & 3030 & 3 \\
\rule[-1ex]{0pt}{3.5ex} D28960 & 2+0.5  &  0.91 & 0.91  & 2690 & 4 \\
\rule[-1ex]{0pt}{3.5ex} E8300  & 0+0.83 &  1.16 & 1.14  & 2330 & 5 \\
\rule[-1ex]{0pt}{3.5ex} E8380  & 1+0.0  &  1.09 & 1.10  & 2620 & 6 \\
\rule[-1ex]{0pt}{3.5ex} E8560  & 1+0.21 &  1.17 & 1.14  & 2610 & 7 \\
\rule[-1ex]{0pt}{3.5ex} P71800 & 0+0.5  &  1.20 & 1.04  & 2160 & 8 \\
\rule[-1ex]{0pt}{3.5ex} P73200 & 1+0.0  &  1.03 & 0.99  & 3130 & 9 \\
\rule[-1ex]{0pt}{3.5ex} P73600 & 1+0.5  &  1.49 & 1.12  & 1930 & 10 \\
\rule[-1ex]{0pt}{3.5ex} P74200 & 2+0.0  &  1.04 & 1.11  & 3060 & 11 \\
\rule[-1ex]{0pt}{3.5ex} P74600 & 2+0.5  &  1.17 & 1.02  & 2200 & 12 \\
\rule[-1ex]{0pt}{3.5ex} P75800 & 3+0.0  &  1.13 & 1.06  & 3060 & 13 \\
\rule[-1ex]{0pt}{3.5ex} P76200 & 3+0.5  &  1.13 & 0.96  & 2270 & 14 \\
\rule[-1ex]{0pt}{3.5ex} P77000 & 4+0.0  &  1.17 & 1.14  & 2870 & 15 \\
\rule[-1ex]{0pt}{3.5ex} M96400 & 0+0.5  &  0.93 & 0.92  & 2310 & 16 \\
\rule[-1ex]{0pt}{3.5ex} M97600 & 1+0.0  &  1.19 & 1.15  & 2750 & 17 \\
\rule[-1ex]{0pt}{3.5ex} M97800 & 1+0.5  &  0.88 & 0.90  & 2460 & 18 \\
\rule[-1ex]{0pt}{3.5ex} M98800 & 2+0.0  &  1.23 & 1.19  & 2650 & 19 \\
\rule[-1ex]{0pt}{3.5ex} O64210 & 0+0.5  &  1.12 & 1.09  & 2050 & 20 \\
\rule[-1ex]{0pt}{3.5ex} O64530 & 0+0.8  &  0.93 & 0.95  & 2150 & 21 \\
\rule[-1ex]{0pt}{3.5ex} O64700 & 1+0.0  &  1.05 & 1.01  & 2310 & 22 \\
\hline
\end{tabular}
}
\end{center}
\end{table}

{\it Monochromatic radius $R_{\lambda}$ and Rosseland radius $R$}.
We use the conventional stellar radius definition where
the monochromatic radius
$R_{\lambda}$ of a star
at wavelength $\lambda$ is given by the distance from the star's
center at which the optical depth equals unity ($\tau_{\lambda}$\,=\,1).
In analogy, the photospheric stellar radius $R$ (Rosseland radius) is given by the
distance from the star's center at which the Rosseland optical depth
equals unity ($\tau_{\rm Ross}$\,=\,1).
This radius has the advantage of agreeing well (see Table 6 and the discussion
in HSW98 for deviations sometimes occurring in very cool stars)
with measurable near-infrared continuum radii
and with the standard boundary radius of pulsation models with $T_{\rm eff} \propto
(L/R^2)^{1/4}$.

{\it Stellar filter radius $R_{\rm f}$}.
For the K-band filter used for the observations we have calculated the theoretical CLVs
corresponding to the above mentioned five Mira star models at different phases and cycles.
The stellar radius
for filter transmission ${\rm f}_{\lambda}$
is the intensity and filter weighted radius
$R_{\rm f} = \int R_{\lambda}\,I_{\lambda}\,{\rm f}_{\lambda}\,d\lambda\,/\,\int I_{\lambda}\,{\rm f}_{\lambda}\,d\lambda$,
which we call stellar filter radius $R_{\rm f}$
after the definition of Scholz \& Takeda$^{19}$.
% Scholz \& Takeda (1987).
In this equation
$R_{\lambda}$ denotes the above monochromatic $\tau_{\lambda}$\,=\,1 radius,
$I_{\lambda}$ the central intensity spectrum and ${\rm f}_{\lambda}$ the transmission of the filter.

{\it Observed angular stellar K-band radius $R_{\rm K,m}^a$ and observed angular Rosseland radius $R_{\rm m}^a$}.
The observed angular stellar K-band radii $R_{\rm K, m}^a$
of the observed Miras
corresponding to the
model-phase combinations m (see Table~\ref{tab:link}), were derived by least-squares fits between the
measured visibilities and the visibilities of the corresponding theoretical CLVs.
Additionally, the angular Rosseland radii $R_{\rm m}^a$
were derived from the obtained
stellar K-band radii $R_{\rm K, m}^a$ and the theoretical ratios $R_{\rm m}$/$R_{\rm K,m}$ from
Table~\ref{tab:link} (Table~\ref{tab:link} provides theoretical
$R$ and $R_{\rm K}$ values for each model-phase combination m).
In the following subsections we apply CLVs predicted from all five models at phases both
near our observations and, for comparison, also at other phases.
\subsection{Effective temperature}
Effective temperatures of each observed Mira star were derived from its angular
Rosseland radii $R_{\rm m}^a$
and
its bolometric flux using the relation
\begin{equation}
T_{\rm eff} = 2341~{\rm K} \times (F_{\rm bol}/\phi^2)^{1/4}
\end{equation}
where $F_{\rm bol}$ is the apparent bolometric flux in units of 10$^{-8}$\,erg\,cm$^{-2}$\,s$^{-1}$
and $\phi$\,=\,2$\times$$R_{\rm m}^a$ is the angular Rosseland diameter in mas.
The bolometric flux was derived from JHKLM-band observations carried out twelve days after the visibility
observations.
For cool stars such as LPVs, where most
of the luminosity is emitted at near-infrared wavelengths, a convenient approximation
for calculating bolometric magnitudes is to use a blackbody function
to interpolate between photometric observations in the J, H, K, L and M bands.
For estimating the bolometric flux we used JHKLM photometric measurements
which were carried out
with the 1.25~m telescope at the Crimean station of
the Sternberg Astronomical Institute
in Moscow twelve days after our visibility observations.

Fig.~\ref{fig:Teff} shows a comparison of the measured and theoretical effective temperatures.
Table~\ref{tab:Fboltab} lists the measured bolometric flux and
the average measured effective temperature for each of the five
observed Mira stars.
For R~Aql we also derived effective temperatures
for the best-fitting D and P models:

\vspace{-0.3cm}
\begin{table} [h]
%\begin{center}
\begin{tabular}{lllll}
\rule[-1ex]{0pt}{3.5ex} Measured effective temperature of R~Aql: & 3007$\pm$155\,K & (D model$^*$); & 3109$\pm$168\,K & (P model$^{**}$) \\
\rule[-1ex]{0pt}{3.5ex} Theoretical D and P model effective temperature: & 3025\,K & (D model$^*$); & 3030\,K & (P model$^{**}$) \\
\end{tabular}
%\end{center}
\end{table}
\vspace{-0.5cm}
\noindent
($^*$ average over phases 1.0, 2.0; $^{**}$ average over phases 1.0, 2.0, 3.0, 4.0)

\begin{table} [ht]
\caption[]{Observational data and measured effective temperatures.}
\label{tab:Fboltab}
\begin{center}
{\small
\begin{tabular}{|c|c|c|c|c|l|}
\hline
\rule[-1ex]{0pt}{3.5ex} Star & Date & $\Phi_{\rm vis}$ & $K$   & F$_{\rm bol}$              & $T_{\rm eff}$ \\
\rule[-1ex]{0pt}{3.5ex}      &      &                  & [mag] & [$10^{-8}$\,erg/cm$^2$\,s] & [K] \\
\hline
\rule[-1ex]{0pt}{3.5ex} X Oph & 99 May 27 & 0.71 & -0.83 & 320.6$\pm$50.1             & 2926$\pm$152$^{**}$ \\
\rule[-1ex]{0pt}{3.5ex} R Aql & 99 May 28 & 0.17 & -0.86 & 351.2$\pm$52.7             & 3072$\pm$161$^{*}$ \\
\rule[-1ex]{0pt}{3.5ex} RU Her & 99 May 21 & 0.07 & -0.11 & 159.8$\pm$24.0             & 2959$\pm$152$^{*}$ \\
\rule[-1ex]{0pt}{3.5ex} R Ser & 99 May 21 & 0.28 & 0.02 &  170.2$\pm$25.5             & 3112$\pm$160$^{**}$ \\
\rule[-1ex]{0pt}{3.5ex} V CrB & 99 May 27 &  0.07 & 0.96 &   52.8$\pm$8.0              & 2325$\pm$122$^{*}$ \\
\hline
\end{tabular}
}
\end{center}
\end{table}
\vspace{-0.8cm}
\noindent
{\small
($^*$ average over all near-maximum model-phase combinations m since the phase of the observation was near-maximum) \\
($^{**}$ average over all model-phase combinations m since models with phases close to the observation do not exist)
}

\subsection{Linear radii}
\begin{figure}
\begin{center}
\psfig{figure=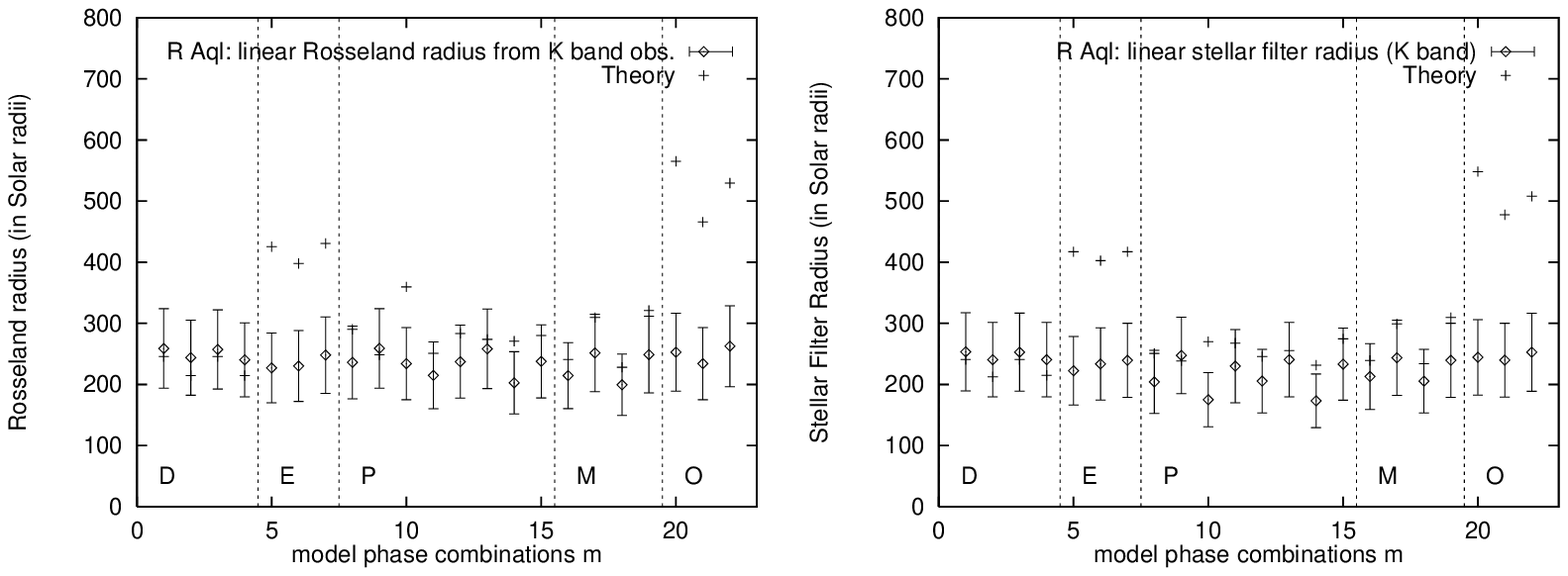,height=6.5cm}
\caption[linradii]{ \label{fig:linradii} Comparison of measured R~Aql radii and theoretical model radii:
(left) linear Rosseland radii $R_{\rm m}$ and (right) linear stellar K-band radii $R_{\rm K,m}$
for all 22 model-phase combinations m.
Table \ref{tab:link} gives the link between the abscissa values (model-phase combinations m)
and the models and their phases.
}
\end{center}
\end{figure}

We have derived linear stellar K-band radii $R_{\rm K,m}$ and Rosseland
radii $R_{\rm m}$ of R~Aql from the measured angular stellar K-band radii $R_{\rm K, m}^a$
and Rosseland radii $R_{\rm m}^a$ by using the R~Aql HIPPARCOS parallax
of 4.73$\pm$1.19\,mas$^{20}$.
The HIPPARCOS parallaxes of the
other four observed Miras have too large errors
for estimating useful linear radii.
Fig.~\ref{fig:linradii}
shows the obtained linear Rosseland radii $R_{\rm m}$ and stellar K-band radii $R_{\rm K, m}$
of R~Aql for all model-phase combinations m.
The theoretical Rosseland radii of the D, M
and P
model series at all available near-maximum phases are close
(within the error bars)
to the measured Rosseland radii of R~Aql.
The theoretical Rosseland radii of the first-overtone models E and O are clearly
too large compared with
measured Rosseland radii.
The same conclusions are also valid for the linear stellar filter radii
$R_{\rm K}$ (Fig.~\ref{fig:linradii}).

If we calculate average R~Aql radii by averaging the radii derived with all available {\it near-maximum}
D model CLVs (i.e., m\,=\,1, 3) and/or {\it near-maximum} P model CLVs (i.e., m\,=\,9, 11, 13, 15)
we obtain:
\newpage
\vspace{-0.3cm}
\begin{table} [h]
%\begin{center}
\begin{tabular}{lll}
\rule[-1ex]{0pt}{3.5ex} Average theoretical D model Rosseland radius: & 246\,R$_{\odot}$ & \\
\vspace{0.2cm}
\rule[-1ex]{0pt}{3.5ex} Average measured D model R~Aql Rosseland radius: & 258\,R$_{\odot}$\,$\pm$65\,R$_{\odot}$ & (obtained with m\,=\,1 and 3) \\
%\rule[-1ex]{0pt}{3.5ex}  & (average over m values 17 and 19) \\
\rule[-1ex]{0pt}{3.5ex} Average theoretical P model Rosseland radius: & 263\,R$_{\odot}$ & \\
\vspace{0.2cm}
\rule[-1ex]{0pt}{3.5ex} Average measured P model R~Aql Rosseland radius: & 242\,R$_{\odot}$\,$\pm$61\,R$_{\odot}$ & (obtained with m\,=\, 9, 11, 13 and 15) \\
%\rule[-1ex]{0pt}{3.5ex} & (average over m values 9, 11, 13 and 15)  \\
\rule[-1ex]{0pt}{3.5ex} Average theoretical D/P model Rosseland radius: & 255\,R$_{\odot}$ & \\
\rule[-1ex]{0pt}{3.5ex} Average measured D/P model R~Aql Rosseland radius: & 250\,R$_{\odot}$\,$\pm$63\,R$_{\odot}$ & (obtained with m\,=\,1, 3, 9, 11, 13, 15, \\
\rule[-1ex]{0pt}{3.5ex} & & 17, 19; D and P models) \\
%\rule[-1ex]{0pt}{3.5ex} & (average over m values 9, 11, 13, 15, 17, 19; M and P models)
\end{tabular}
%\end{center}
\end{table}

\subsection{Pulsation mode}
Adopting the above phase-averaged (over models D and P at maximum phases)
linear Rosseland radius of 250\,R$_{\odot}$\,$\pm$63\,R$_{\odot}$ for R~Aql,
we find for the pulsation constant
$Q = P\,(M/M_{\odot})^{1/2}\,(R/R_{\odot})^{-3/2}$
a value of $Q$=0.072$\pm$0.027 for a 1\,M$_{\odot}$-Mira with period $P$=284\,days.
This $Q$ value agrees within the 1$\sigma$ error with the theoretical value ($Q$=0.088)
for fundamental pulsation mode
for 1\,M$_{\odot}$-AGB stars with a period of $\sim$\,284 days$^{21}$.
% (Fox\,\&\,Wood 1982).
The corresponding $Q$ value of first overtone pulsation mode
is $Q$=0.049. Note, however, that no direct measurement of a Mira mass exists
and that a 20\% uncertainty of $M$ would for example result in a 10\% uncertainty of $Q$.
%
%\newpage
\section{DISCUSSION}
We derived the following angular uniform-disk diameters $\phi$ of five Mira stars from
K-band visibility measurements with the 38\,m baseline of the IOTA interferometer
and the FLUOR beam combiner:
\vspace{-0.3cm}
\begin{table} [h]
%\begin{center}
\begin{tabular}{ll}
\rule[-1ex]{0pt}{3.5ex} X~Oph: & $\phi$\,=\,11.7\,mas\,$\pm$0.3\,mas  \\
\rule[-1ex]{0pt}{3.5ex} R~Aql: & $\phi$\,=\,10.9\,mas\,$\pm$0.3\,mas \\
\rule[-1ex]{0pt}{3.5ex} RU~Her: & $\phi$\,=\,8.4\,mas\,$\pm$0.2\,mas \\
\rule[-1ex]{0pt}{3.5ex} R~Ser: & $\phi$\,=\,8.1\,mas\,$\pm$0.2\,mas \\
\rule[-1ex]{0pt}{3.5ex} V~CrB: & $\phi$\,=\,7.9\,mas\,$\pm$0.2\,mas \\
\end{tabular}
%\end{center}
\end{table}
\vspace{-0.5cm}

\noindent
The following effective temperatures were obtained from photometric JHKLM observations
and the derived angular Rosseland radii:
\newpage
\vspace{-0.3cm}
\begin{table} [h]
%\begin{center}
\begin{tabular}{lll}
\rule[-1ex]{0pt}{3.5ex} X~Oph &($\Phi_{\rm vis}$=0.71): & 2926\,K\,$\pm$152\,K \\
\rule[-1ex]{0pt}{3.5ex} R~Aql &($\Phi_{\rm vis}$=0.17): & 3072\,K\,$\pm$161\,K \\
\rule[-1ex]{0pt}{3.5ex} RU~Her &($\Phi_{\rm vis}$=0.07): & 2959\,K\,$\pm$152\,K \\
\rule[-1ex]{0pt}{3.5ex} R~Ser &($\Phi_{\rm vis}$=0.28): & 3112\,K\,$\pm$160\,K \\
\rule[-1ex]{0pt}{3.5ex} V~CrB &($\Phi_{\rm vis}$=0.07): & 2325\,K\,$\pm$122\,K \\
\end{tabular}
%\end{center}
\end{table}
\vspace{-0.5cm}

\noindent
Previous interferometric K-band observations of some of our target stars
(R~ Aql, X~Oph, R~Ser) were carried out
by van Belle et al.$^8$
% van Belle et al. (1996)
at similar phases.
Their derived uniform-disk diameters (R~Aql; $\Phi_{\rm vis}$\,=\,0.90: 10.76$\pm$0.61\,mas,
X~Oph; $\Phi_{\rm vis}$\,=\,0.75: 12.30$\pm$0.66\,mas,
R~Ser; $\Phi_{\rm vis}$\,=\,0.32: 8.56$\pm$0.58\,mas) are in good agreement
with our observations (within the error bars).
Their measured effective temperatures (R~Aql: 3189$\pm$147\,K, X~Oph: 3041$\pm$160\,K,
R~Ser: 2804$\pm$144\,K) are also in good agreement with our results, with the exception of R~Ser.

\noindent
For R~Aql a good HIPPARCOS parallax (4.73$\pm$1.19\,mas) is available and it is therefore possible to
compare measured linear Rosseland and stellar K-band radii with the theoretical radii of the BSW96 and
HSW98 models.
The measured radii were derived by fitting theoretical (BSW96, HSW98)
center-to-limb intensity
variations to the visibility data.
In the following table we compare measured and theoretical values:

\vspace{-0.3cm}
\begin{table} [h]
%\begin{center}
\begin{tabular}{lllll}
\rule[-1ex]{0pt}{3.5ex} Measured linear R~Aql Rosseland radii $R_{\rm m}$: & 258$\pm$65\,R$_{\odot}$ & (D model$^*$); & 242$\pm$61\,R$_{\odot}$ & (P model$^{**}$)  \\
\rule[-1ex]{0pt}{3.5ex} Theoretical linear Rosseland radii $R_{\rm m}$: & 246\,R$_{\odot}$ & (D model$^*$); & 263\,R$_{\odot}$ & (P model$^{**}$) \\
\rule[-1ex]{0pt}{3.5ex} Measured linear R~Aql stellar K-band radii $R_{\rm K,m}$: & 253$\pm$64\,R$_{\odot}$ & (D model$^*$); & 238$\pm$61\,R$_{\odot}$ & (P model$^{**}$) \\
\rule[-1ex]{0pt}{3.5ex} Theoretical linear stellar K-band radii $R_{\rm K,m}$: & 241\,R$_{\odot}$ & (D model$^*$); & 259\,R$_{\odot}$ & (P model$^{**}$) \\
\rule[-1ex]{0pt}{3.5ex} Measured R~Aql effective temperature: & 3007$\pm$155\,K & (D model$^*$); & 3109$\pm$168\,K & (P model$^{**}$) \\
\rule[-1ex]{0pt}{3.5ex} Theoretical effective temperature: & 3025\,K & (D model$^*$); & 3030\,K & (P model$^{**}$) \\
\end{tabular}
%\end{center}
\end{table}
\vspace{-0.5cm}
\noindent
($^*$ average over phases 1.0, 2.0; $^{**}$ average over phases 1.0, 2.0, 3.0, 4.0)

\noindent
The comparison suggests that R~Aql is well represented by the fundamental mode D and P model (BSW96, HSW98).
The measured Rosseland radius of $R$\,=\,250$\pm$63\,R$_{\odot}$ (average of the derived values from D and
P model CLVs; corresponding theoretical D/P model Rosseland radius\,=\,255\,R$_{\odot}$)
places R~Aql among the fundamental mode pulsators in the
period-radius relation which also is in agreement with Ref. 8.
% van Belle et al. (1996).
Note, however, that observations in more filters than just one continuum filter may be necessary
for safely distinguishing a well-fitting model from an accidental match
(cf. Ref. 10).
% (cf. Hofmann et al. 2000).
%
%
\section{REFERENCES}
1. M.S. Bessell, M. Scholz, P.R. Wood, {\it A\&A} {\bf 307}, pp. 481, 1996 (BSW96) \\
2. K.-H. Hofmann, M. Scholz, P.R. Wood, {\it A\&A} {\bf 339}, pp. 846, 1998 (HSW98) \\
3. D. Bonneau, A. Labeyrie, {\it ApJ} {\bf 181}, pp. L1, 1973 \\
4. M. Karovska, P. Nisenson, C. Papaliolios, R.P. Boyle, {\it ApJ} {\bf 374}, pp. L51, 1991 \\
5. A. Quirrenbach, D. Mozurkewich, J.T. Armstrong, et al., {\it A\&A} {\bf 259}, pp. L19, 1992 \\
6. C.A. Haniff, M. Scholz, P.G. Tuthill, {\it MNRAS} {\bf 276}, pp. 640, 1995 \\
7. G. Weigelt, Y. Balega, K.-H. Hofmann, M. Scholz, {\it A\&A} {\bf 316}, pp. L21, 1996 \\
8. G.T. Van Belle, H.M. Dyck, J.A. Benson, M.G. Lacasse, {\it AJ} {\bf 112}, pp. 2147, 1996 \\
9. G. Perrin, V. Coud$\acute{\rm e}$ du Foresto, S.T. Ridgway, et al., {\it A\&A} {\bf 345}, pp. 221, 1999 \\
10. K.-H. Hofmann, Y. Balega, M. Scholz, G. Weigelt, {\it A\&A} {\bf 353}, pp. 1016, 2000 \\
11. T. Watanabe, K. Kodaira, {\it PASJ} {\bf 31}, pp. 61, 1979 \\
12. M. Scholz, {\it A\&A} {\bf 145}, pp. 251, 1985 \\
13. M.S. Bessell, J.M. Brett, M. Scholz, P.R. Wood, {\it A\&A} {\bf 213}, pp. 209, 1989 \\
14. N.P. Carleton, W.A. Traub, M.G. Lacasse, et al., {\it Proc. SPIE} {\bf 2200}, pp. 152, 1994 \\
15. W.A. Traub, et al., {\it Proc. SPIE} {\bf 3350}, pp. 848, 1998 \\
16. V. Coud$\acute{\rm e}$ du Foresto, S.T. Ridgway, J.-M. Mariotti, {\it A\&AS} {\bf 121}, pp. 379, 1997 \\
17. G. Perrin, V. Coud$\acute{\rm e}$ du Foresto, S.T. Ridgway, et al. {\it A\&A} {\bf 331}, pp. 619, 1998 \\
18. H.M. Dyck, J.A. Benson, G.T. Van Belle, S.T. Ridgway, {\it AJ} {\bf 111(4)}, pp. 1705, 1996 \\
19. M. Scholz, Y. Takeda, {\it A\&A} {\bf 186}, pp. 200, 1987 (erratum: 196, 342) \\
20. F. Van Leeuwen, M.W. Feast, P.A. Whitelock, B. Yudin, {\it MNRAS} {\bf 287}, pp. 955, 1997 \\
21. M.W. Fox, P.R. Wood, {\it ApJ} {\bf 259}, pp. 198, 1982 \\

  \end{document}